\documentclass[aps,pre,groupedaddress,11pt,tightenlines,longbibliography]{revtex4-1}
\usepackage[hmargin=1.5in]{geometry}
\usepackage[T1]{fontenc}

\pdfoutput=1
\usepackage[dvipsnames,rgb,dvips,hyperref]{xcolor}
\usepackage{graphicx}
\usepackage{float}
\usepackage{overpic}
\usepackage{amsmath,amssymb,amsfonts,bbm}
\usepackage{hyperref}
\definecolor{linkblue}{HTML}{2200CC}
\hypersetup{
    colorlinks,
    linkcolor=linkblue,
    citecolor=linkblue,
    urlcolor=linkblue
}

\makeatletter
\def\amsbb{\use@mathgroup \M@U \symAMSb}
\makeatother
\usepackage[bbgreekl]{mathbbol}
\usepackage{mathrsfs}

\newcommand{\lc}{\ensuremath{\varepsilon}}

\newcommand{\Tr}{\ensuremath{\textrm{Tr}\,}}

\newcommand\nn{\nonumber}
\newcommand\ve[1]{\boldsymbol{#1}}

\newcommand{\ensemble}[1]{\ensuremath{\overline{#1}}}
\newcommand{\rd}{\ensuremath{\textrm{d}}}
\newcommand{\tr}{\ensuremath{^{\rm T}}}

\newcommand{\Wi}{\ensuremath{\textrm{Wi}}}

\newcommand{\eqnlab}[1]{\label{eqn:#1}}
\newcommand{\figlab}[1]{\label{fig:#1}}

\newcommand{\eqnref}[1]{(\ref{eqn:#1})}

\newcommand{\Eqnref}[1]{Eq.~(\ref{eqn:#1})}
\newcommand{\Figref}[1]{Fig.~\ref{fig:#1}}

\newcommand{\applab}[1]{\label{app:#1}}

\newcommand{\Appref}[1]{Appendix~\ref{app:#1}}

\usepackage{scalerel,stackengine}
\stackMath
\newcommand\reallywidehat[1]{%
\savestack{\tmpbox}{\stretchto{%
  \scaleto{%
    \scalerel*[\widthof{\ensuremath{#1}}]{\kern-.6pt\bigwedge\kern-.6pt}%
    {\rule[-\textheight/2]{1ex}{\textheight}}
  }{\textheight}%
}{0.5ex}}%
\stackon[1pt]{#1}{\tmpbox}%
}

\begin{document}

\title{The Einstein viscosity with fluid elasticity}

\author{Jonas Einarsson}
\author{Mengfei Yang}
\author{Eric S.G. Shaqfeh}
\affiliation{Department of Chemical Engineering, Stanford University}
\date{\today}

\begin{abstract}
We give the first correction to the suspension viscosity due to fluid elasticity for a dilute suspension of spheres in a viscoelastic medium. Our perturbation theory is valid to $O(\phi\Wi^2)$ in the Weissenberg number $\Wi=\dot\gamma \lambda$, where $\dot\gamma$ is the typical magnitude of the suspension velocity gradient, and $\lambda$ is the relaxation time of the viscoelastic fluid. 
For shear flow we find that the suspension shear-thickens due to elastic stretching in strain {\lq}hot spots{\rq} near the particle, despite the fact that the stress inside the particles decreases relative to the Newtonian case. We thus argue that it is crucial to correctly model the extensional rheology of the suspending medium to predict the shear rheology of the suspension.
For uniaxial extensional flow we correct existing results at $O(\phi\Wi)$, and find dramatic strain-rate thickening at $O(\phi\Wi^2)$. We validate our theory with fully resolved numerical simulations.
\end{abstract}

\maketitle

\section{Introduction}
The rheology of complex suspensions is of fundamental importance in many fields of science and engineering. 
For most applications it not desirable, or practical, to resolve the details at the scale of the polymers or particles in the suspension. A fundamental problem in rheology is the development of methods to coarse-grain the microscopic fluid dynamics and thus create a useful macroscopic continuum description of the stress-strain response of a complex suspension. This is a difficult problem at the intersection of fluid dynamics and statistical mechanics.

A cornerstone of our theoretical understanding is so-called dilute suspension rheology \cite{batchelor1970}. This approximation describes the stresses that arise due to isolated particles in a suspending medium, and particle-particle interactions are neglected. Mathematically, it arises at first order in a perturbation theory in the volume fraction $\phi$ of particles in the suspension \cite{hinch1977}.

\citet{einstein1906,einstein1911} first devised an early form of dilute suspension rheology in his doctoral thesis. He showed that the bulk shear viscosity $\eta$ of a suspension of rigid, inertia-free, and neutrally buoyant spheres in a Newtonian fluid is 
\begin{align}
\eta = (1+2.5\phi + ...)\mu\,,\eqnlab{einstein}
\end{align}
where $\mu$ is the viscosity of the suspending fluid in the absence of any particles. The suspension viscosity increases because the particles resist deformation. Their internal stresses increase, which results in an $O(\phi)$ increase in the suspension viscosity. In a seminal paper, \citet{batchelor1970} described how to generalize Einstein's calculation to compute the complete suspension stress tensor. He termed the increased particle stress the {\lq}stresslet{\rq} contribution, because of its relation to the symmetric first moment of surface tractions over the particle surface.

When the suspending medium is viscoelastic, for example by the addition of polymers, there are two mechanisms that change the resulting suspension stress. First, the stresslet contribution may change, because the surface tractions change. Second, in contrast to the Newtonian case, there is additional stress in the fluid phase due to the polymers stretching in the flow gradients induced by the particle. We call this latter contribution the {\lq}particle induced fluid stress{\rq} \cite{yang2016}.

Experiments with suspensions of spherical particles in viscoelastic fluids show shear-thickening at low ($<10\%$) volume fractions of particles \cite{zarraga2001,scirocco2005,dai2014}. In contrast, Newtonian suspensions shear-thin or thicken only at rather high volume fractions $\phi > 30$-$40\%$, because of particle interactions \cite{wagner2009}. These observations indicate that dilute suspension rheology may be useful to understand the mechanism of shear thickening of viscoelastic suspensions.

The first deviation from a Newtonian fluid for a viscoelastic fluid medium is described by the Second-order fluid \cite{larson2013}, which is an approximation in slow flows or, equivalently, fast relaxation of the elastic fluid. It is valid to linear order in the Weissenberg number $\Wi=\dot\gamma \lambda$, where $\dot\gamma$ is the typical magnitude of the fluid velocity gradient, and $\lambda$ is the relaxation time of the viscoelastic fluid. \citet{koch2006,koch2008} first calculated the correct rheology of a dilute suspension of spheres in a Second-order fluid, after several earlier attempts with conflicting results \cite{kaloni1983,sun1984,mifflin1985,li1989,greco2005,greco2007a,greco2007,housiadas2009}. Recently, \citet{yang2016} could discriminate between the different theoretical calculations by fully resolved numerical simulations. But in the Second-order fluid limit there is no correction to the shear viscosity. While there are Second-order fluid corrections to the normal stress differences due to fluid elasticity \cite{koch2006,koch2008}, Einstein's result \eqnref{einstein} remains the leading order correction to the suspension viscosity.

\citet{yang2016} and \citet{koch2016} independently studied suspension stress in shear flow by numerical simulation and a semi-analytical theory, respectively. They both found that as $\Wi$ is increased from vanishingly small values, the stresslet contribution to the dilute suspension shear viscosity decreases, but the particle induced fluid stress increases. Thus both found that the net effect is shear thickening of viscosity and the first normal stress.

In this paper we analytically calculate the suspension stress for any linear motion of the suspension, by a perturbation theory to $O(\phi \Wi^2)$. As a particular case we find the correction to the Einstein viscosity \eqnref{einstein} due to fluid elasticity. Our calculation reveals how shear-thickening arises from strain {\lq}hot-spots{\rq} in the disturbance flow around particles. In addition we analyze the stress in extensional flow, which is a fundamental rheological flow, and important in applications where, for example, a suspension is injected into a mold.



\section{Theory}
\subsection{Dilute suspension rheology}
We consider an inertia-free suspension of rigid spherical particles of radius $a$ in a viscoelastic medium. We take the macroscopic flow to be a linear flow $\ve U=\ve E\cdot\ve x+\ve O\cdot\ve x$, where the rate of strain $\ve E$ is symmetric and the vorticity tensor $\ve O$ is anti-symmetric. The linear flow includes both the simple shear and extensional flows that are fundamental to rheology. In the following we first explain the general averaging procedure, we then give the details of the microscopic problem, and how to calculate the average.

The macroscopic stress at a point $\ve x$ in the suspension is an ensemble average of the microscopic stress $\ve \sigma(\ve x)$, taken over all possible configurations of the suspension \cite{batchelor1970,hinch1977}. 
In a dilute suspension it is unlikely that two particles are close enough to interact, and for low volume fraction $\phi$ the ensemble average is approximated by \cite{hinch1977}
\begin{align}
   \ensemble{\ve \sigma(\ve x)}  = \int \ve \sigma(\ve x\, |\, \ve y)P(\ve y)\,\rd\ve y + O(\phi^2)\,. \eqnlab{singleaverage}
\end{align}
Here $\ve \sigma(\ve x\,|\,\ve y)$ is the stress at $\ve x$ conditioned on the presence of a sphere centered at $\ve y$, and $P(\ve y)$ is the probability to find a sphere at $\ve y$. \Eqnref{singleaverage} embodies the fact that we expect to find only a single particle within a volume $V\sim\phi^{-1}$. In a spatially homogenous suspension the probability $P(\ve y)=\phi/V_p$ is uniform, where $V_p$ is the volume of a sphere, and the disturbance fields due to the presence of a particle depend only on $\ve r = \ve x-\ve y$. 

It follows that the correct microscopic problem to solve, at this order, is that of a single particle centered at $\ve r=0$ in an asymptotically large volume of radius $R\sim\phi^{-1/3}$. Outside this volume we expect to find additional particles, and the assumptions that led to \Eqnref{singleaverage} are invalid. The problem formulation must be closed by a far field boundary condition for $|\ve r|\sim R$ that ensures a self-consistent theory, which we introduce after giving the equations of motion.

The ensemble average \eqnref{singleaverage} thus reduces to 
\begin{align}
  \ensemble{\ve \sigma } = \frac{\phi}{V_p}\int_V \ve \sigma(\ve r)\,\rd\ve r=
  \frac{\phi}{V_p}\int_{V_p} \ve \sigma(\ve r)\,\rd\ve r+\frac{\phi}{V_p}\int_{V_f} \ve \sigma(\ve r)\,\rd\ve r
  \,, \eqnlab{volumeaverage}
\end{align}
where $V$ is the volume of the domain, including $V_p$, and the fluid volume $V_f = V-V_p$.
The stress $\ve\sigma$ inside the particle is unknown, but via the divergence theorem and continuity of stress it is given by the stresslet of a freely suspended particle \cite{batchelor1970}
\begin{align}
  \frac{\phi}{V_p}\int_{V_p} \ve \sigma(\ve r)\,\rd\ve r = \frac{\phi}{V_p}\int_{S_p} \ve r (\ve \sigma\cdot\ve n)\rd S = \frac{\phi}{V_p}\ve S\,.\eqnlab{stresslet}
\end{align}

\subsection{Governing equations}
The microscopic flow $\ve u$ is governed by the momentum equation and the incompressibility condition
\begin{align}
  \nabla \cdot \ve \sigma = 0\,, \quad \nabla \cdot \ve u = 0\,,\eqnlab{eqnmotion}
\end{align}
where $\ve \sigma$ is the stress tensor field. In the following all variables are non-dimensionalized: $\ve r'=\ve r/a$, $t'=\dot\gamma t$, $\ve u'=\ve u/(\dot\gamma a)$, $\ve \sigma' = \ve\sigma/(\mu \dot\gamma)$, where $\mu$ is the shear viscosity at $\Wi=0$ and $\dot\gamma=\sqrt{2\Tr \ve E\ve E}$. We subsequently drop the primes from the notation. We use the Oldroyd-B constitutive model for the stress in a viscoelastic fluid. It represents a thermal bath of entropic springs, modeling polymers, that are transported and stretched by the fluid. The model captures the rheology of an elastic fluid without shear-thinning. The steady Oldroyd-B equations read \cite{larson2013}
\begin{align}
  &\ve \sigma = -p\ve \delta + 2(1-\mu_r)\ve e + \mu_r \ve \Pi\,,\nn\\
  &\ve \Pi + \Wi [(\ve u\cdot\nabla)\ve \Pi - \ve a\cdot\ve \Pi - \ve \Pi \cdot\ve a\tr] = \ve a + \ve a\tr\,.
  \eqnlab{constitutive}
\end{align}
Here $p$ is the pressure, $\mu_r\ve \Pi$ is the stress due to the elastic polymers, and $\ve a$ is the flow gradient tensor with elements $a_{ij}=\partial u_i/\partial r_j$. The strain tensor is defined as $\ve e=(\ve a+\ve a\tr)/2$, and the vorticity tensor $\ve o=(\ve a-\ve a\tr)/2$, so that $\ve a= \ve e + \ve o$. The constitutive model has two parameters: the relaxation time $\lambda$ that appears in the Weissenberg number $\Wi=\dot\gamma\lambda$, and the relative concentration of polymers given by the ratio $\mu_r=\mu_p/(\mu_s+\mu_p)$ of the solvent ($\mu_s$) and polymer ($\mu_p$) contributions to the shear viscosity $\mu=\mu_s+\mu_p$ at $\Wi=0$.
The center-of-mass velocity $\ve v$ and the angular velocity $\ve \omega$ of the particle are determined by the condition that it is force- and torque-free.

\subsection{Boundary conditions}
The boundary condition on the surface of the sphere is the usual no-slip condition. The asymptotic far field boundary condition when $|\ve r|\sim R$ is determined by two self-consistency conditions. First, the suspension flow is linear by assumption, so the microscopic model must satisfy
\begin{align}
\ensemble{ \nabla \ve u } =\ve E+\ve O\,.  
\end{align}
This implies that the flow field settles down to its mean value as $|\ve r|\sim\phi^{-1/3}$. Any correction due to the presence of another particle is of higher order in $\phi$ \cite{hinch1977}. Second, the ensemble average 
\begin{align}
\ensemble{ (\ve u\cdot\nabla)\ve \Pi }=0\,  
\end{align}
in a homogenous suspension \cite{koch2016,rallison2012} (see also \Appref{bc}.) This implies that also the stress field settles down to its mean value as $|\ve r|\sim\phi^{-1/3}$. In summary, 
\begin{align}
  \ve u \sim \ensemble{ \ve u}\,,~~ \ve \sigma \sim \ensemble{ \ve \sigma} \,,~~  |\ve r|\sim \phi^{-1/3}\,. \eqnlab{bc}
\end{align}
This argument is a mean field theory in the sense that we \emph{assume} a macroscopic average field, and then determine the microscopic model to satisfy this assumption upon averaging. We demonstrate in \Appref{bc} that these conditions indeed give a consistent theory. 

Viscous flow problems in unbounded domains are often complicated by the slow algebraic decay of the disturbance fields, giving unphysical or divergent integrals. The self-consistency condition is in effect a regularization of those integrals, and our argument is akin to the regularization used to calculate the $O(\phi^2)$ rheology in a Newtonian suspension \cite{obrien1979}. They considered an unbounded fluid domain, but instead invoked the asymptotic properties of $\ve u$ and $\ve \sigma$ when evaluating the integrals corresponding to our \Eqnref{volumeaverage}. We show in \Appref{bc} that this procedure gives the same result as ours. 

If one chooses to approximate the microscopic problem with an unbounded flow without regularization, it appears that the volume averaged stress is not equal to the ensemble averaged stress, even in a statistically homogenous suspension \cite{koch2006,rallison2012,koch2016}. This apparent ergodicity breaking has led to confusion regarding the correct averaging procedure, where some terms required ensemble averaging, whereas others could be volume averaged \cite{greco2005,koch2006,rallison2012,koch2016}. We remove this ambiguity by correctly imposing the mean field conditions.

\subsection{Averaging}
It follows from \Eqnref{volumeaverage}, \eqnref{stresslet}, and the Oldroyd-B constitutive equation \eqnref{constitutive} that
\begin{align}
 \ensemble{ \ve \sigma }
  = -\langle p\rangle_F\ve\delta \!+\! 2\ve E + \frac{\phi}{V_p}\ve S + \mu_r \Wi \big\langle\ve a\ve \Pi + \ve \Pi\ve a\tr\big\rangle_F \,,\eqnlab{firstaverage}
\end{align}
where
\begin{align}
  \langle \ve a \rangle_F \equiv \frac{\phi}{V_p}\int_{V_f}\ve a\,\rd V\,.
\end{align}
is short-hand for the integral over the fluid volume in \Eqnref{volumeaverage}.
In \Eqnref{firstaverage} we used that the integrals $\langle \ve e\rangle_F = \ve E$, and $\langle(\ve u\cdot\nabla)\ve \Pi\rangle_F = 0$ given the boundary conditions \eqnref{bc}. In the following we omit any isotropic terms in the average stress, because they do not contribute to the suspension rheology. We denote the symmetric part of any tensor $\widehat{\ve a}\equiv(\ve a + \ve a\tr)/2$.
With the perturbation ansatz $\ve \Pi = \ve \Pi^{(0)} + \Wi\, \ve \Pi^{(1)} + ...$ we have
\begin{align}
  &\ensemble{ \ve \sigma} = 2\ve E + \frac{\phi}{V_p}\ve S+ 
  \mu_r \Wi[ 2\langle\reallywidehat{\ve a\cdot \ve a}\rangle_F + 2\langle\ve a\cdot \ve a\tr\rangle_F] \nn\\
  &\,\,+\mu_r \Wi^2[2\langle\reallywidehat{\ve a\cdot\ve a\cdot\ve a}\rangle_F + 6\langle\reallywidehat{\ve a\cdot\ve a\cdot\ve a\tr}\rangle_F - 
  4\big\langle\reallywidehat{\ve a \cdot[(\ve u\cdot\nabla)\ve e]}\big\rangle_F ]\,,\eqnlab{finalvolumeavg}
\end{align}
where the stresslet $\ve S$ must be evaluated to $O(\Wi^2)$, the integrals in the first bracket must be evaluted to $O(\Wi)$, and those in the second bracket to $O(1)$.

The integrals over the quadratic and cubic terms are calculated by splitting $\ve a = \ensemble{ \ve a } + \ve a'$, and noting that $\langle \ve a' \rangle_F=\phi \ve E$. This is because the average strain in the fluid phase is $(1+\phi)\ve E$, to compensate for the fact that there is no strain inside the particle. The remaining integrals of the type $\langle \ve a'\cdot\ve a'\rangle_F$, $\langle \ve a'\cdot\ve a'\cdot\ve a'\rangle_F$, and $\langle\ve a' \cdot[(\ve u\cdot\nabla)\ve e]\rangle_F$ are evaluated using the flow solutions that satisfy the boundary conditions \eqnref{bc} to $O(1)$ in $\phi$, which gives an accurate result to $O(\phi)$.

We evaluate the stresslet via the Lorentz reciprocal theorem \cite{kim1991,einarsson2017}
\begin{align}
  \ve S
  = 
  \frac{20\pi}{3}\ve E 
  +\! \int_{S_p}\!\!\! \reallywidehat{\ve r (\ve\sigma^{E} \cdot \ve n)} \,\rd S
  +\! \int_V\!\!\! \ve M\tr\cdot \nabla \cdot \ve\sigma^{E}\rd V.\eqnlab{rt_stresslet}
\end{align}
Here $\ve\sigma^E=\mu_r(\ve\Pi - 2\ve e)$ is the non-linear part of the stress tensor, and $\ve M$ is the rank three tensor such that $\ve M : \ve E$ is the Stokes solution for a sphere in a otherwise quiescent fluid, but with a strain flow $\ve E\cdot\ve r$ on the surface. See \Appref{rt} for a detailed derivation of \Eqnref{rt_stresslet}. 

We must know the flow and stress fields associated with the microscopic problem to $O(\Wi)$ to evaluate the average \eqnref{finalvolumeavg} and the stresslet \eqnref{rt_stresslet} to $O(\phi\Wi^2)$. The required flow and stress fields are given by a regular perturbation theory in $\Wi$, 
 \begin{align}
   \ve u = \ve u^{(0)}+\Wi\,\ve u^{(1)}+...\,,\quad \ve \Pi = \ve \Pi^{(0)} + \Wi\, \ve \Pi^{(1)} + ...\,,\nn\\
   p = p^{(0)}+\Wi\,p^{(1)}+...\,,\quad \ve \omega = \ve \omega^{(0)} + \Wi\, \ve \omega^{(1)} + ...\,.
 \end{align}
We follow the method described in Ref.~\citenum{einarsson2017} to calculate these solutions. We perform the necessary algebra and integrations to evaluate Eqns.~\eqnref{finalvolumeavg} and \eqnref{rt_stresslet} using the \emph{Matte} software in Mathematica \cite{matte}.

\section{Numerical methods}

The numerical evaluations of the bulk stress is obtained by using a massively parallel flow solver based on an unstructured finite volume formulation.  Details for the solver can be found in \cite{yang2016} and the references therein.  In those previous studies, the full Navier-Stokes equation was solved and the only modification we have made in this study is to remove the convective term such that the solver is an unsteady Stokes solver.  At steady state, i.e. when the stresslet components and the particle-induced fluid stress components display a relative change of less than $0.001$ over a characteristic time scale, the solutions discretely satisfy the governing equations \eqnref{eqnmotion} and \eqnref{constitutive}. 

The computation domain is a cubic box with a sphere at the center.  For the shear flow results, we use the same boundary conditions as reported in Ref.~\citenum{yang2016}.  At the sphere surface, the velocity corresponds to the solid body rotation of the sphere with angular velocity $\omega$, which is determined by iterating until the non-dimensionalized torque is less than 0.01.   Equal and opposite velocities are applied on two walls of the computation box to drive the shear flow and periodic boundaries are applied on the remaining four walls.  Note that we don't need to apply boundary conditions for the polymer stress in this case because the two walls and sphere surface have zero mass flux.  For the extensional flow results, we impose zero velocity on the sphere surface and $\ve u=\ve E\cdot\ve r$ on the computational box.  In this case, since there is mass flux through the computation box, we also impose the ensemble averaged value of the polymer stress at the computation box to match the boundary conditions \eqnref{bc}.

We have tested for mesh, time, and domain size convergence.  Results change by less than 1.5\% when we decrease the mesh size by a factor of 2, decrease the time step by a factor of 5, and increase the domain size by a factor of 2.  We use a computation box that is 12 times the particle size.  The mesh consists of tetrahedral elements that are finer on the sphere surface (0.03 particle diameters) and coarser on the boundary of the computation box (0.5 particle diameters) for a total of 2.5 million control volumes.  


To extract the dependence of the stresslet on the value of the Weissenberg number we must subtract the $O(1)$ Newtonian contribution. Despite the numerical results for the viscosities being within $1.5$\% of the theoretical Newtonian values at small values of $\Wi$, subtracting the theoretical value is not precise enough to make a power law on a log-log scale. We therefore fit a curve $\eta=a+b\Wi^2$ (or $a+b\Wi+c\Wi^2$ for extensional flow) to the three data-points with lowest $\Wi$, and use $a$ as the value to subtract. The values of $a$ are within $1.5$\% of the theoretical Newtonian value for all data-sets shown.

\section{Results}
We give our results in rounded decimal form for a concise and useful presentation. The exact expressions may be found in \Appref{exact}.

\begin{figure}
  \includegraphics{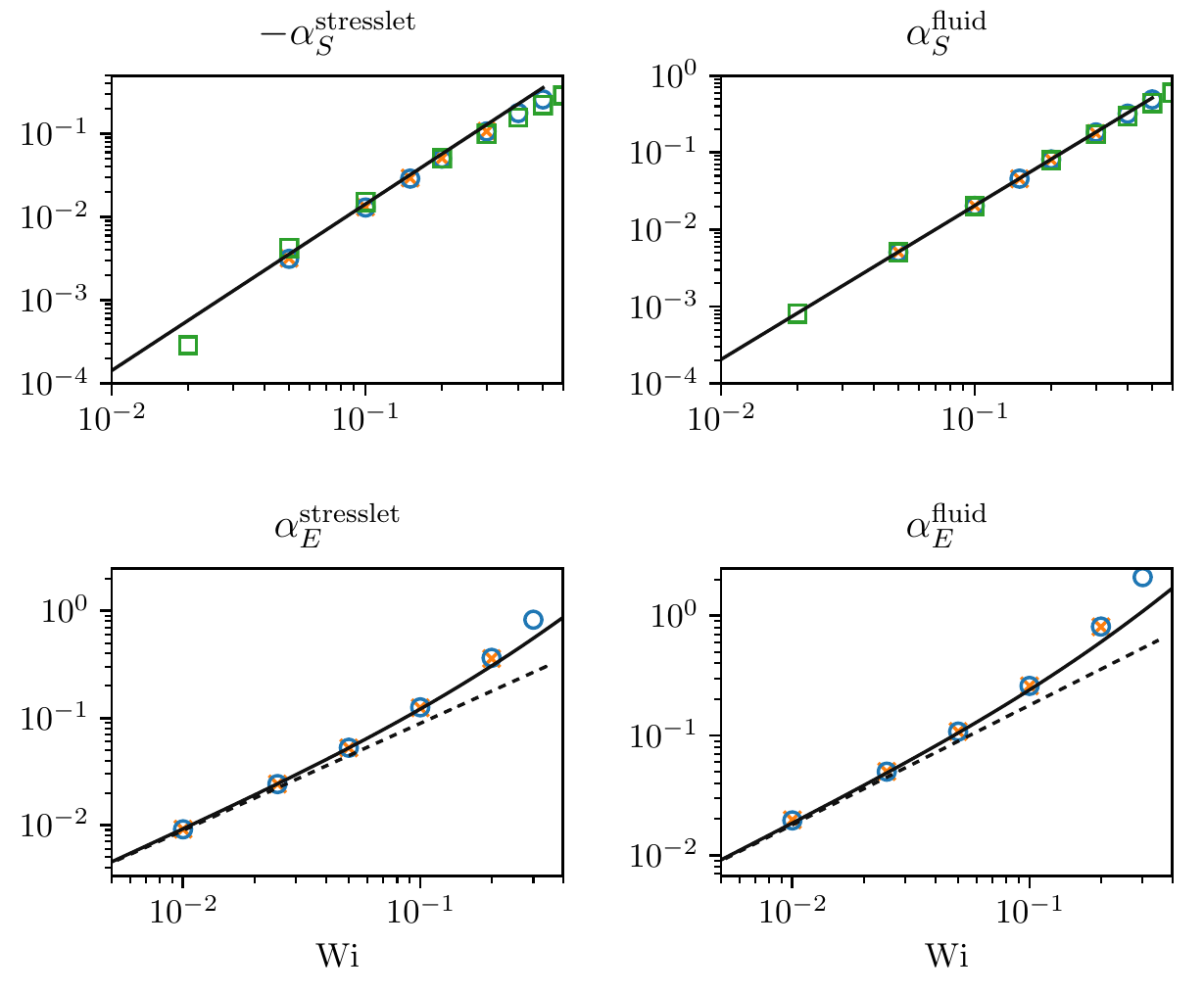}\\
  \caption{\figlab{numerical} Comparison of the theoretical results (lines) to fully resolved numerical simulations (markers) of the particle contribution to the suspension viscosities, as function of $\Wi$. 
  Top row shows suspension shear viscosity, bottom extensional viscosity in uniaxial strain.
  Left panels show stresslet contribution, right panels particle induced fluid stress. In all panels markers are numerical results for $\mu_r=0.68$ (red crosses), $\mu_r=0.5$ (green squares), and $\mu_r=0.01$ (blue circles). The solid lines show the $O(\Wi^2)$ theoretical results from \Eqnref{etasplit} and \Eqnref{etaEsplit}. The dotted lines show the $O(\Wi)$ terms of \Eqnref{etaEsplit}.
  }
\end{figure}
\begin{figure}
  \includegraphics{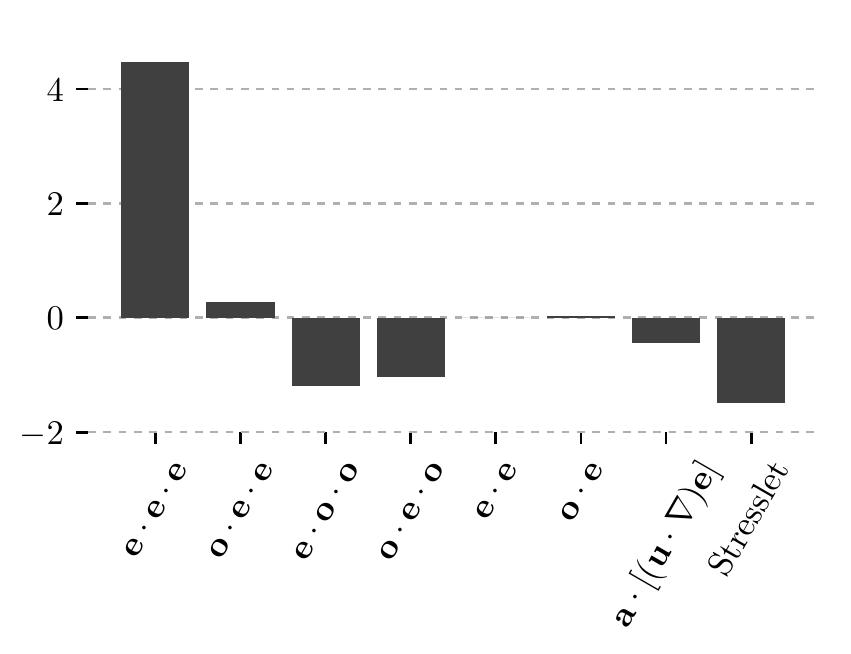}
  \caption{\figlab{terms} Contributions to the $O(\phi\Wi^2)$ suspension shear viscosity for $\mu_r=1$. The leftmost six bars represent contributions from gradient correlations in \Eqnref{finalvolumeavg} with $\ve a = \ve e + \ve o$. The seventh bar represents the effect of the convective term. These seven bars add up to $\alpha^{\mathrm{fluid}}_S$.}
\end{figure}

\subsection{Shear viscosity}

When the suspension flow is a simple shear flow $\ve U = y \hat{\ve x}$, we find that the shear viscosity $\eta_S = \ensemble{ \sigma_{xy}}$ is
\begin{align}
\eta_S = 1+2.5\phi+ \left(0.62 - 0.03\mu_r\right)\phi\mu_r\Wi^2\,.\eqnlab{mainresultShear}
\end{align}
The suspension is shear-thickening. The thickening is a consequence of two competing mechanisms. To show this we consider the components of the $O(\phi\Wi^2)$ contribution: $\eta_S=1+2.5\phi + \phi\mu_r (\alpha_S^{\mathrm{stresslet}}+\alpha_S^{\mathrm{fluid}})$. 
\begin{align}
  \alpha_S^{\mathrm{stresslet}}&=-1.43\Wi^2 - 0.06\mu_r\Wi^2\,,\nn\\
  \alpha_S^{\mathrm{fluid}}&=2.05\Wi^2+0.03\mu_r\Wi^2\,.\eqnlab{etasplit}
\end{align}
There is a negative, shear-thinning, contribution from the stresslet component, meaning that the stress in the sphere decreases due to the elasticity of the suspending fluid. 
The angular velocity also decreases at $O(\Wi^2)$ \cite{housiadas2011a}: $\omega_z = -1/2 + \mu_r\Wi^2/4$. Thus both the symmetric and anti-symmetric moments of the surface tractions decrease at this order. 
On the other hand, the average polymer stress in the fluid phase gives a strong positive contribution to the shear viscosity. Our interpretation is that the polymers {\lq}absorb{\rq} some of the stress that would otherwise have gone to surface traction on the particle. But that results in stretched polymers in the fluid around the particle that, on average, incur an even larger extra stress in the suspension. The net result is that the suspension shear-thickens. We compare our analytical result to fully resolved numerical simulations in \Figref{numerical} (top row). 

We further elucidate the mechanism of thickening by splitting the gradient $\ve a = \ve e + \ve o$ in \Eqnref{finalvolumeavg}. The resulting strain and vorticity correlations represent contributions from different flow types to the $O(\phi\Wi^2)$ viscosity correction. We present the contributions from all integrals, as well as the stresslet, in \Figref{terms}, for $\mu_r=1$. It is clear that the shear-thickening is due to the particle induced fluid stress in regions of flow around the particle with a strong straining component, represented by the integral $\langle \ve e \cdot\ve e \cdot\ve e\rangle_F$. The gradients in the regions fore and aft of the particle are strain-dominated, in the sense that the gradient tensor has three real eigenvalues. In the shear direction, however, the gradients are mixed strain and rotation. Regions of mixed flow diminish the effect of the strain, represented by the negative contributions of for example $\langle \ve e \cdot\ve o \cdot\ve o\rangle_F$. Nevertheless, the net result is thickening due to the regions of strain-dominated flow. We demonstrate this correlation in \Figref{field_subplots} which displays the local flow type side by side with the local contribution to the $O(\phi\Wi^2)$  particle induced fluid stress. The flow type is represented by the eigenvalue discriminant of the gradient tensor \cite{chong1990}, and the stress contribution is the integrand in the second bracket of \Eqnref{finalvolumeavg}.

\begin{figure}
  \includegraphics{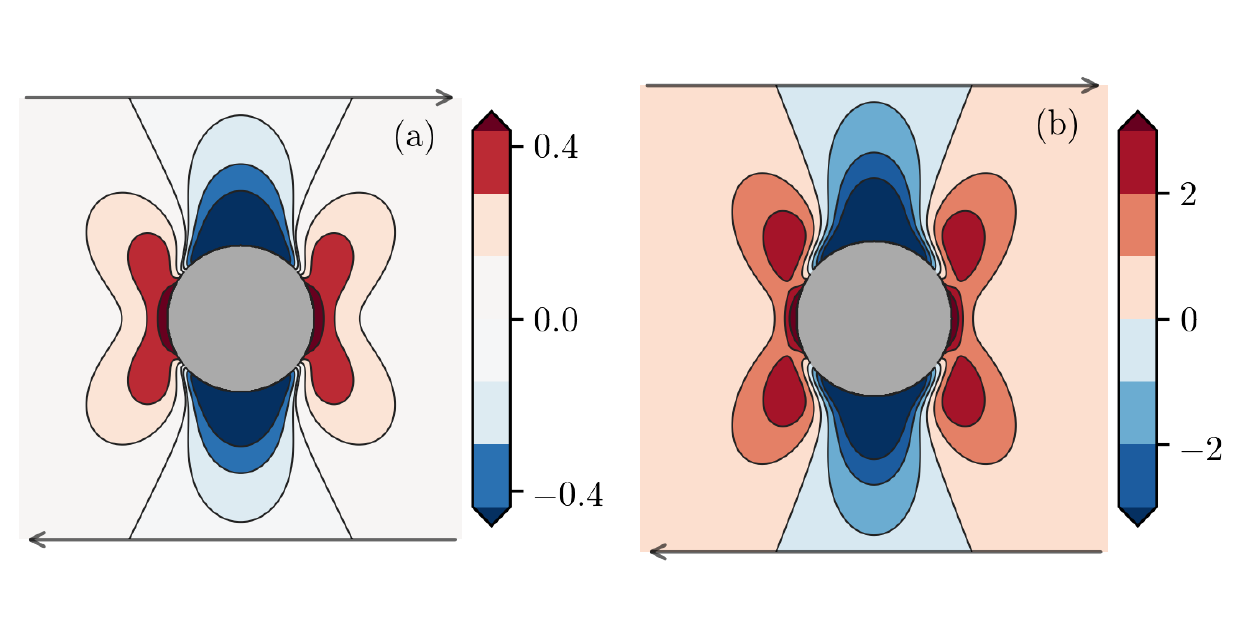}
  \caption{\figlab{field_subplots} Section in flow-shear plane of two fields: (a) Flow type, represented by the gradient eigenvalue discriminant $\Delta=(\Tr \ve a^2)^3-6(\Tr\ve a^3)^2$. Positive values of $\Delta$ imply strain-dominated flow where $\ve a$ has three real eigenvalues. (b) Integrand for $O(\phi\Wi^2)$ contribution to $\alpha_S^{\mathrm{fluid}}$ (the $xy$-component of the second bracket in \Eqnref{finalvolumeavg}). The shear-thickening contributions to the suspension shear viscosity comes from regions of strain-dominated flow.}
\end{figure}
\begin{figure}
  \includegraphics{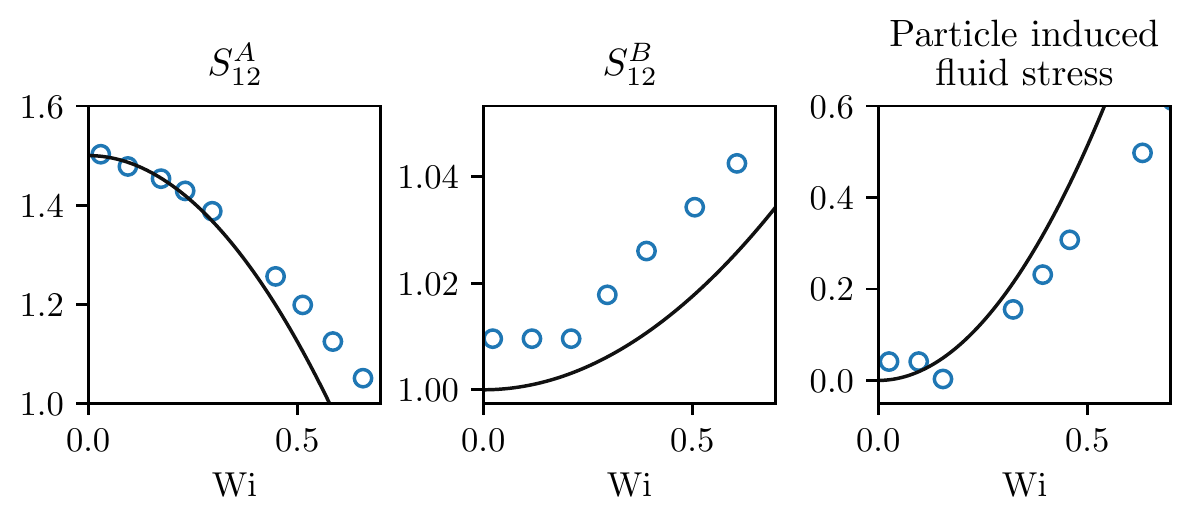}
  \caption{\figlab{koch_compare}Comparison of theory (solid lines) to data (markers) presented in Figs.~1 and 2 in \citet{koch2016}. Left panel: Surface integral component of stresslet in reciprocal theorem, \Eqnref{sa}. Center panel: Volume integral component of stresslet in reciprocal theorem, \Eqnref{sb}. Right panel: Particle induced fluid stress.}
\end{figure}

We compare our theory to the semi-analytical results by \citet{koch2016} for small values of $\Wi$.
We digitized the data from their Figs.~1 and 2 and compare to the relevant expressions from our theory in \Figref{koch_compare}. The first two panels show the two contributions to the stresslet,
\begin{align}
  \ve S^A &=
  \! \int_{S_p}\!\!\! \reallywidehat{\ve r (\ve\sigma^{E} \cdot \ve n)} \,\rd S \eqnlab{sa}\\
  \ve S^B &=
  \! \int_V\!\!\! \ve M\tr\cdot \nabla \cdot \ve\sigma^{E}\rd V.\eqnlab{sb}
\end{align}
The third panel shows the particle induced fluid stress.
While the digitized data is too crude to make a quantitative comparison to a $\Wi^2$ power law, we see that it is in good qualitative agreement with the theory.

The calculation by \citet{koch2016} is valid to linear order in $\mu_r$. This approximation neglects the feedback mechanism that the elastic stresses modify the flow velocity field, which in turn generates new elastic stresses, which is represented by terms proportional to $\mu_r^2$ in our calculation. Our result shows that their model is a good approximation to $O(\phi\mu_r\Wi^2)$, because the terms of $O(\phi \mu_r^2 \Wi^2)$ are very small.

Finally, we note that the normal stress differences are unaffected at $O(\phi\Wi^2)$, and at $O(\phi\Wi)$ we recover the known results \cite{koch2016,yang2016}.

\subsection{Extensional viscosity}

For uniaxial elongational flow $\ve U=x \hat{\ve x}-\frac{1}{2}y\hat{\ve y}-\frac{1}{2}z\hat{\ve z}$ we find that the elongational viscosity $\eta_E = (\ensemble{\sigma_{xx}}-\ensemble{\sigma_{yy}})/3$ is
\begin{align}
 &\eta_E = 1 + \mu_r\Wi + 3\mu_r\Wi^2 + 2.5\phi
  \nn\\
 &~+ \phi \mu_r \left(2.68\Wi+9.36 \Wi^2 - 0.1\mu_r\Wi^2\right)\eqnlab{mainresultE}
\end{align}
In this flow the leading correction to the viscosity is $O(\phi\Wi)$. Our result at this order differs from that of \citet{greco2005} because they mistakenly evaluated the average over the fluid volume in \Eqnref{volumeaverage} to $O(1)$ instead of $O(\phi)$ \cite{greco2007,rallison2012}. They therefore report only the stresslet contribution, which is in agreement with our result for the stresslet. We separate the contributions from the stresslet and the particle induced fluid stress as $\eta_E=1 + \mu_r\Wi + 3\mu_r\Wi^2 + 2.5\phi + \phi\mu_r (\alpha_S^{\mathrm{stresslet}}+\alpha_S^{\mathrm{fluid}})$: 
\begin{align}
  \alpha_E^{\mathrm{stresslet}}&=0.89\Wi+3.2\Wi^2-0.17\mu_r\Wi^2\,,\nn\\
  \alpha_E^{\mathrm{fluid}}&=1.79\Wi+6.16\Wi^2+0.08\mu_r \Wi^2\,.\eqnlab{etaEsplit}
\end{align}
Both stresslet and particle induced fluid stress thicken the suspension, and the dominant contribution arises from the fluid stress around the particle. We note that the coefficient of $\Wi^2$ is so large that the second order result is important already for $\Wi\approx0.1$ (\Figref{numerical}, bottom row.)

\section{Conclusions}
We have calculated the stress in a dilute suspension of rigid spherical particles in an elastic fluid to $O(\phi\Wi^2)$. This revealed the first effects of elasticity on the suspension viscosity. When shearing the suspension we found that the stress in the particle phase decreases relative to the Newtonian case, but in return the stress in the elastic fluid surrounding the spheres increases. The net result is that the suspension is shear thickening. The main contribution to this thickening is the enhanced polymer stress in the straining regions of the fluid surrounding the spheres. This is significant because it implies that it is crucially important to correctly model the \emph{extensional} rheology of the suspending fluid, in order to model the \emph{shear} rheology of the suspension.

The available measurements of shear rheology in viscoelastic particle suspensions display shear thickening \cite{zarraga2001,scirocco2005,dai2014}. It is, however, not possible to extract an $\Wi^2$ trend from their data for quantitative comparison. Further, the suspending fluids in the experiments were characterized by their shear rheology alone. We now believe that a successful quantitative comparison to experiment requires modeling the extensional rheology of the suspending medium. Thus, we argue that experimental measurements correlating the suspension rheology to the extensional rheology of the suspending medium will be valuable to further our understanding of complex suspensions at all values of $\Wi$.

The contributions proportional to $\mu_r^2 \Wi^2$ represent the elastic stress disturbing the flow field, which in turn generates new elastic stresses. This effect is very small at this order of approximation, and therefore the $O(\mu_r)$ semi-analytical theory for low polymer concentration \cite{koch2016} is a good approximation at least up to $O(\phi\Wi^2)$.

We also gave the first correction to the extensional rheology of the suspension. In this case both the stresslet and particle induced fluid stresses contribute to significant thickening of the suspension with increasing strain rate. A recent experiment \cite{dai2017} shows strain-rate thickening in a rather dense suspension ($\phi=40\%$). We are, however, not aware of any experimental measurements of the extensional viscosity of a dilute suspension of spheres in an elastic fluid.

%
\appendix

\section{Exact results}\applab{exact}

In the main text we present our results in rounded decimal form for convenience. For reference, the exact results are for shear flow:
\begin{align}
  \eta_S=1+\frac{5}{2}\phi + \phi\mu_r (\alpha_S^{\mathrm{stresslet}}+\alpha_S^{\mathrm{fluid}})
\end{align}
\begin{align}
  \alpha_S^{\mathrm{stresslet}}&=-\frac{83645}{58344}\Wi^2 - \frac{29405}{504504}\mu_r\Wi^2\,,\nn\\
  \alpha_S^{\mathrm{fluid}}&=\frac{115}{56}\Wi^2+\frac{5}{196}\mu_r\Wi^2\,.
\end{align}
For uniaxial extensional flow:
\begin{align}
  \eta_E=1 + \mu_r\Wi + 3\mu_r\Wi^2 + \frac{5}{2}\phi + \phi\mu_r (\alpha_S^{\mathrm{stresslet}}+\alpha_S^{\mathrm{fluid}})
\end{align}
with
\begin{align}
  \alpha_E^{\mathrm{stresslet}}&=\frac{25}{28}\Wi+\frac{62215}{19448}\Wi^2-\frac{29405 }{168168}\mu_r\Wi^2\,,\nn\\
  \alpha_E^{\mathrm{fluid}}&=\frac{25}{14}\Wi+\frac{345}{56}\Wi^2+\frac{15}{196}\mu_r \Wi^2\,.
\end{align}

\section{On the boundary conditions}\applab{bc}
Here we give some detail to the averaging integrals mentioned in the Letter. For the case of the average velocity gradient, and for the convective term in the polymer stress, we demonstrate that our calculation and the regularization scheme of \citet{obrien1979} are equivalent. We also demonstrate explicitly that considering an unbounded domain without employing any regularization gives an inconsistent result.

In the following the full domain is denoted by $V$, and $S_\infty$ is a far field surface at the asymptotically large distance $R$ (taken to $\phi^{-1/3}$ in our Letter, but to $\infty$ in case of an unbounded domain.) The ensemble average of a quantity $\ve a$ is, in the dilute approximation,
\begin{align}
  \ensemble{ \ve a}= \frac{\phi}{V_p}\int_V\ve a(\ve r)\rd V\,,
\end{align}
of which Eqns.~(3) and (4) in the letter display the case of the stress tensor.

\subsection{Average of the velocity gradient}
The average $\ensemble{ \partial_j u_i }=E_{ij}+O_{ij}$ by assumption in the mean field theory. The direct calculation is, by the divergence theorem,
\begin{align}
 \ensemble{ \partial_j u_i }= \frac{\phi}{V_p}\int_V \partial_j u_i \rd V = \frac{\phi}{V_p}\int_{S_\infty} \! u_i n_j \rd S\,.
\end{align}
Our boundary condition requires $u_i \sim (E_{ik}+O_{ik})r_k$, when $r\sim\phi^{-1/3}$:
\begin{align}
   \frac{\phi}{V_p}\int_{S_\infty} \! (E_{ik}+O_{ik})r_k n_j \,\rd S = E_{ij}+O_{ij} \,,
\end{align}
where we used that $\phi/V_p=1/V$, and
\begin{align}
  \int_S n_i r_j \rd S = \int_V\partial_i r_j\rd V= V\delta_{ij}\,
\end{align}
for any closed domain $V$ bounded by the surface $S$.

The method of \citet{obrien1979} amounts to arguing that in the differential surface element $\rd S$, $\ve r$ samples a large enough region to achieve statistical stationarity for $\ve u$, and therefore $\ve u$ may be replaced by its statistical average in the integral. This obviously yields the same integral as our boundary condition.

Finally, should one assume an unbounded domain approximation for $\ve u$, the leading order flow field is
\begin{align}
  u_i &= (E_{ik}+O_{ik})r_k - \frac{5}{2r^5}E_{kl}r_ir_kr_l + O(\frac{1}{r^4})\,,
\end{align}
one finds
\begin{align}
  \frac{\phi}{V_p}\int_{S_\infty} \! u_i n_j \rd S = E_{ij}+O_{ij}-\frac{5\phi}{2V_p}E_{kl}\int_{S_\infty} \! n_in_kn_l n_j \rd \Omega = E_{ij}+O_{ij}-\phi E_{ij}\,.
\end{align}
Here we took $S_\infty$ to be a sphere at distance $R$, as $R\to\infty$. The integral is convergent as $R\to\infty$ because $\ve u$ decays at the same rate as the surface element $\rd S$ grows. However, the result does not satisfy the mean field theory to $O(\phi)$.

\subsection{Average of the convective term}
The general argument for why $\ensemble{ (\ve u\cdot\nabla)\ve \Pi}=0$ is as follows. The ensemble average commutes with the gradient operator, and $\ve u$ is divergence-free, so
\begin{align}
  \ensemble{ (\ve u\cdot\nabla)\ve \Pi} = \ensemble{\nabla\cdot \ve u\ve \Pi-(\nabla\cdot \ve u)\ve \Pi} =  \ensemble{\nabla\cdot\ve u\ve \Pi}= \nabla\cdot\ensemble{\ve u\ve \Pi}\,.
\end{align}
Split $\ve u=\ensemble{\ve u}+\ve u'$ and $\ve \Pi=\ensemble{\ve\Pi}+\ve \Pi'$, where primed quantities denote fluctuations around the mean. Then
\begin{align}
  \nabla\cdot\ensemble{\ve u\ve \Pi}&=\nabla\cdot\ensemble{\ve u}\ensemble{\ve \Pi}+\nabla\cdot\ensemble{\ve u'\ve \Pi'}\nn\\
&=(\ensemble{\ve u}\cdot\nabla)\ensemble{\ve \Pi}+\nabla\cdot\ensemble{\ve u'\ve \Pi'}=0\,.
\end{align}
Because $\ve \Pi$ and $\ve u'$ are statistically homogenous in the spatial variables by assumption, the gradient of their averages vanish. This argument was given earlier by \citet{koch2006} and \citet{rallison2012}.

The direct calculation from the microscopic problem proceeds by the divergence theorem:
\begin{align}
  \ensemble{ u_k\partial_k\Pi_{ij}} = \frac{\phi}{V_p}\int_V u_k\partial_k\Pi_{ij} \rd V = \frac{\phi}{V_p}\int_{S_\infty} u_kn_k\Pi_{ij} \rd S\,.
\end{align}
The polymer stress vanishes inside the particle, and $u_kn_k=0$ on the particle surface.
Our boundary condition requires $u_k \sim (E_{kl}+O_{kl})r_l$, and $\Pi_{ij}\sim\ensemble{\Pi_{ij}}$, when $r\sim\phi^{-1/3}$:
\begin{align}
  \frac{\phi}{V_p}\int_{S_\infty} u_kn_k\Pi_{ij} \rd S\sim\frac{\phi}{V_p}\int_{S_\infty} (E_{kl}+O_{kl})r_ln_k\ensemble{\Pi_{ij}} \rd S\,.
\end{align}
The tensor $\ve \Pi$ is spatially homogenous by assumption, so $\ensemble{\Pi_{ij}}$ is a constant. The integral vanishes because $E_{kk}=O_{kk}=0$.

The argument of \citet{obrien1979} leads to the same conclusion. In that case we would argue that both $\ve u$ and $\ve \Pi$ are statistically stationary within the surface element $\rd S$, and may therefore be replaced by their ensemble averages in the integral.

Finally, let us evaluate the volume averaged convective term using the leading order flow solutions in an unbounded domain without any regularization.
The far field asymptotes of those fields read
\begin{align}
  u_kn_k &\sim (E_{kl}+O_{kl})r_ln_k + O(\frac{1}{r^2})\,,\nn\\
  \Pi_{ij} &\sim 2E_{ij} - 5\frac{E_{ik}r_kr_j + E_{jk}r_kr_i+\delta_{ij}E_{kl}r_kr_l}{r^5} + 25\frac{E_{kl}r_ir_jr_kr_l}{r^7}+O(\frac{1}{r^5})\,.
\end{align}
After a bit of algebra it follows that
\begin{align}
  \frac{\phi}{V_p}\int_{S_\infty}u_kn_k\Pi_{ij} \rd S \sim \phi\left(\frac{12}{7} E_{ik}E_{kj}-\frac{4}{7}\delta_{ij} E_{kl}E_{lk}\right)\,,\quad R\to\infty\,.\eqnlab{convfinal}
\end{align}
The integrand decays as $1/r^2$, which gives a convergent, but inconsistent result for the integral. We note that the volume average may also be evaluated directly as a volume integral. In that case a scaling analysis indicates that the integral is divergent, because the integrand decays as $1/r^3$. But that contribution vanishes identically upon integration of the angular variables, and the remaining radial integrals in fact decay as $1/r^6$ or faster, yielding the result in \Eqnref{convfinal}.

\section{Reciprocal theorem}\applab{rt}

Let $\ve u$ and $\ve \sigma^N$ denote the flow field and Newtonian stress tensor of a flow that is governed by the equation of motion
\begin{align}
  \partial_j \sigma^N_{ij}=f_i\,,\quad \partial_i u_i = 0\,.
\end{align}
The general form of Lorentz reciprocal theorem is \cite{kim1991}
\begin{align}
  \int_S \tilde u_i \sigma^N_{ij} n_j\,\rd S - \int_V \tilde u_i \partial_j \sigma^N_{ij}\,\rd V
  = \int_S  u_i \tilde\sigma_{ij} n_j\,\rd S - \int_V u_i \partial_j \tilde \sigma_{ij}\,\rd V\,,
\end{align}
where $\tilde{\ve u}$ and $\tilde{\ve \sigma}$ are the flow and stress fields of an auxiliary Newtonian flow problem in the same geometry but with different boundary conditions.
The volume integral is over a volume of fluid, and the surface integrals are to be taken over all boundaries of $V$ with surface normals pointing out from $V$.

Take the auxiliary problem to be the Stokes flow governed by
\begin{align}
  \partial_j \tilde\sigma_{ij}&=0\,,\quad \partial_i \tilde u_i = 0\,\nn\\
  \tilde u_i &= \tilde E_{ik}r_k\quad \textrm{on particle surface,}\nn\\
  \tilde u_i&\to0\,,\quad r\to\infty\,.
\end{align}
Then
\begin{align}
  \int_S \tilde u_i \sigma^N_{ij} n_j\,\rd S - \int_V \tilde u_i f_i\,\rd V
  = \int_S  u_i \tilde\sigma_{ij} n_j\,\rd S \,.
\end{align}
We apply this theorem to our problem by noting that
\begin{align}
  \ve \sigma = \ve \sigma^N + \ve \sigma^E\,,
\end{align}
where
\begin{align}
  \ve \sigma^N &= -p\ve \delta + 2\ve e\,,\nn\\
  \ve \sigma^E &= \mu_r(\ve \Pi - 2\ve e)\,.
\end{align}
Thus
\begin{align}
  \int_S \tilde u_i \sigma_{ij} n_j\,\rd S 
  = \int_S \tilde u_i \sigma^E_{ij} n_j\,\rd S + \int_S  u_i \tilde\sigma_{ij} n_j\,\rd S
  - \int_V \tilde u_i \partial_j\sigma^E_{ij}\,\rd V \,.
\end{align}
The surface integral on the left hand side is related to the stresslet because $\tilde u_i = \tilde E_{ik}r_k$ on the surface of the particle. The volume integral is well defined because $\tilde u_i\to0\,,~r\to\infty$. The auxiliary flow is thus the disturbance flow around a sphere in a straining flow, and it decays as $1/r^2$ as $r\to\infty$. It is given by
\begin{align}
  \tilde u_i &= \frac{1}{r^5}\tilde E_{ij}r_j + \frac{5}{2}\left(\frac{1}{r^5}-\frac{1}{r^7}\right)\tilde E_{jk}r_ir_jr_k\,,\nn\\
  \tilde p &= \frac{5}{r^5}\tilde E_{jk}r_jr_k\,.
\end{align}

Split the surface integrals into the integral over the particle surface $S_p$ and the far-field surface $S_\infty$ that we take to be a sphere at $r=R_\infty$. Using the boundary conditions for both $\ve u$ and $\tilde{\ve u}$ we have
\begin{align}
  \tilde E_{ik}\int_{S_p} r_k \sigma_{ij} n_j\,\rd S 
  &= 
  \int_{S_\infty} \tilde u_i \sigma_{ij} n_j\,\rd S 
  - \int_{S_\infty} \tilde u_i \sigma^E_{ij} n_j\,\rd S 
  + \tilde E_{ik}\int_{S_p} r_k \sigma^E_{ij} n_j\,\rd S \nn\\
  &\quad- \int_{S_\infty}  u_i \tilde\sigma_{ij} n_j\,\rd S
  + \lc_{ilk}\omega_l\int_{S_p}  r_k \tilde\sigma_{ij} n_j\,\rd S
  + \int_V \tilde u_i \partial_j\sigma^E_{ij}\,\rd V \,.
\end{align}
Here $\ve \omega$ is the angular velocity of the particle. Starting from the left have the sought stresslet
\begin{align}
  \tilde E_{ik}\int_{S_p} r_k \sigma_{ij} n_j\,\rd S &= \tilde E_{ik} S_{ik}\,,
\end{align}
Next, 
\begin{align}
  \int_{S_\infty} \tilde u_i \sigma_{ij} n_j\,\rd S- \int_{S_\infty} \tilde u_i \sigma^E_{ij} n_j\,\rd S = \int_{S_\infty} \tilde u_i \sigma^N_{ij} n_j\,\rd S \sim \frac{8\pi}{3}\tilde E_{ik} E_{ik} \,,\quad R_\infty\to\infty\,.
\end{align}
Here we used that $\ve \sigma^N \sim 2\ve E + O(1/r^2)$ and $\tilde{\ve u}\sim \frac{5}{2}\ve r (\ve r\ve r : \tilde{\ve E})/r^5+ O(1/r^4)$, as $r\to\infty$.
Next,
\begin{align}
  \tilde E_{ik}\int_{S_p} r_k \sigma^E_{ij} n_j\,\rd S &= \tilde E_{ik} S^E_{ik}\,,
\end{align}
is the stresslet due to the non-linear polymer surface traction.
Next, 
\begin{align}
  -\int_{S_\infty}  u_i \tilde\sigma_{ij} n_j\,\rd S \sim 4\pi \tilde E_{ik}E_{ik}\,,\quad R_\infty\to\infty\,,
\end{align}
where we used that 
\begin{align}
  \tilde\sigma_{ij} &\sim \frac{5}{r^5}(\tilde E_{ik} r_k r_j+ \tilde E_{jk}r_kr_i)-\frac{25}{r^7}\tilde E_{kl}r_kr_lr_ir_j + O(1/r^5)\,,\nn\\
  u_i &\sim E_{ik}r_k + O(1/r^2)\,,\quad r\to\infty\,.
\end{align}
Next,
\begin{align}
  \lc_{ilk}\omega_l\int_{S_p}  r_k \tilde\sigma_{ij} n_j\,\rd S = 0\,,
\end{align}
because the particle in the auxiliary problem is torque-free.

Taken together we have, as $R_\infty \to \infty$,
\begin{align}
  \tilde E_{ik} S_{ik} 
  &= 
  \frac{20\pi}{3}\tilde E_{ik} E_{ik}
  + \tilde E_{ik} S^E_{ik}
  + \int_V \tilde u_l \partial_j\sigma^E_{lj}\,\rd V \,.
\end{align}
Finally, by renaming the dummy indices and expressing $\tilde u_l = M_{lik}\tilde E_{ik}$ in the volume integral, we eliminate $\tilde{\ve E}$ and arrive at the theorem stated in the main manuscript:
\begin{align}
  \ve S
  = 
  \frac{20\pi}{3}\ve E 
  + \int_{S_p} \widehat{\ve r (\ve\sigma^{E} \cdot \ve n)} \,\rd S
  + \int_V \ve M \tr\cdot \nabla \cdot \ve\sigma^{E}\rd V\,.
\end{align}

\section{Derivation of non-linear stress}\applab{stress}
In this Appendix we employ a shorthand for contraction of rank two tensors, so that $\ve a \cdot \ve \Pi$ is written simply $\ve a\ve\Pi$, in the interest of keeping the expressions brief. There are no outer products of rank two tensors in this manuscript.

Starting from the constitutive equation
\begin{align}
  \ve \Pi + \Wi [(\ve u\cdot\nabla)\ve \Pi - \ve a\ve \Pi - \ve \Pi \ve a\tr] = \ve a + \ve a\tr\,,
\end{align}
insert $\ve \Pi = \ve \Pi^{(0)}+\Wi\,\ve \Pi^{(1)}+\Wi^2\,\ve \Pi^{(2)}$ and compare order by order:
\begin{align}
  \ve \Pi^{(0)} &= \ve a + \ve a\tr\eqnlab{piexp1}\\
  \ve \Pi^{(1)} &= -[(\ve u\cdot\nabla)\ve \Pi^{(0)} - \ve a\ve \Pi^{(0)} - \ve \Pi^{(0)} \ve a\tr]\eqnlab{piexp2}\\
  \ve \Pi^{(2)} &= -[(\ve u\cdot\nabla)\ve \Pi^{(1)} - \ve a\ve \Pi^{(1)} - \ve \Pi^{(1)} \ve a\tr]\eqnlab{piexp3}\,.
\end{align}
Inserting \Eqnref{piexp1} in \Eqnref{piexp2}, and that result into \Eqnref{piexp3}, find
\begin{align}
  \ve \Pi^{(1)} &= \ve a(\ve a + \ve a\tr)+ (\ve a + \ve a\tr) \ve a\tr - (\ve u\cdot\nabla)(\ve a + \ve a\tr) \nn\\
  &= \ve a \ve a + \ve a\tr\ve a\tr + 2\ve a \ve a\tr - (\ve u\cdot\nabla)(\ve a + \ve a\tr)\nn\\
  &= 2\widehat{\ve a \ve a} + 2\ve a \ve a\tr - (\ve u\cdot\nabla)(\ve a + \ve a\tr)\nn\\
  &= 4\widehat{\ve a \ve e} - 2(\ve u\cdot\nabla)\ve e\nn\\
\end{align}
\begin{align}
  \ve \Pi^{(2)} &= 
  \ve a(\ve a \ve a + \ve a\tr\ve a\tr + 2\ve a \ve a\tr - (\ve u\cdot\nabla)(\ve a + \ve a\tr)) 
  + (\ve a \ve a + \ve a\tr\ve a\tr + 2\ve a \ve a\tr - (\ve u\cdot\nabla)(\ve a + \ve a\tr))\ve a\tr \nn\\
  &\qquad- (\ve u\cdot\nabla)(\ve a \ve a + \ve a\tr\ve a\tr + 2\ve a \ve a\tr - (\ve u\cdot\nabla)(\ve a + \ve a\tr)) \nn\\
  &= \ve a \ve a \ve a + 3\ve a \ve a\tr\ve a\tr + 3\ve a\ve a \ve a\tr + \ve a\tr\ve a\tr\ve a\tr - \ve a [(\ve u\cdot\nabla)(\ve a + \ve a\tr)] - [(\ve u\cdot\nabla)(\ve a + \ve a\tr)]\ve a\tr \nn\\
  &\qquad - (\ve u\cdot\nabla)(\ve a \ve a + \ve a\tr\ve a\tr + 2\ve a \ve a\tr) + (\ve u\cdot\nabla)^2(\ve a + \ve a\tr) \nn\\
  &= 2\widehat{\ve a \ve a \ve a} + 6\widehat{\ve a \ve a\ve a\tr} - 2\reallywidehat{\ve a [(\ve u\cdot\nabla)(\ve a + \ve a\tr)]} - 2(\ve u\cdot\nabla)(\widehat{\ve a \ve a} + \ve a \ve a\tr) + (\ve u\cdot\nabla)^2(\ve a + \ve a\tr) \nn\\
  &= 4\widehat{\ve a \ve a \ve e} + 4\widehat{\ve a \ve e\ve a\tr} - 4\reallywidehat{\ve a [(\ve u\cdot\nabla)\ve e]} - 4(\ve u\cdot\nabla)\widehat{\ve a \ve e} + 2(\ve u\cdot\nabla)^2\ve e \,.
\end{align}
Thus
\begin{align}
  \ve \sigma^E &= \mu_r \Wi [4\widehat{\ve a \ve e} - 2(\ve u\cdot\nabla)\ve e] \nn\\
  &\quad + \mu_r \Wi^2[ 4\widehat{\ve a \ve a \ve e} + 4\widehat{\ve a \ve e\ve a\tr} - 4\reallywidehat{\ve a [(\ve u\cdot\nabla)\ve e]} - 4(\ve u\cdot\nabla)\widehat{\ve a \ve e} + 2(\ve u\cdot\nabla)^2\ve e]\,.
\end{align}
To be consistent to $O(\Wi^2)$ the flow fields in the first bracket must be evaluated to $O(\Wi)$, whereas the fields in the second bracket only require the $O(1)$ Stokes flow solution.

\end{document}